\documentclass[prc,aps,showpacs,floatfix,nofootinbib,twocolumn,letterpaper]
{revtex4}
\usepackage{graphicx}
\usepackage{amssymb}
\usepackage{amsmath}
\usepackage{amsfonts}

\def\rb{\rangle}

\def\be{\begin{equation}}
\def\ee{\end{equation}}

\def\adag{a^{\dagger}}
\def\ei{\varepsilon_i}
\def\ev8{{\tt ev8}}

\begin{document}
\title{Number-conserving theory of nuclear pairing gaps:
a global assessment
}
\author{Abhishek Mukherjee,$^{1}$ Y. Alhassid,$^{1}$ and G. F. Bertsch$^{2}$ }
\affiliation{$^{1}$Center for Theoretical Physics, Sloane Physics
Laboratory, Yale University, New Haven, Connecticut 06520, USA\\
$^{2}$Department of Physics and Institute of Nuclear Theory,
Box 351560\\ University of Washington, Seattle, Washington 98915, USA}

\begin{abstract}

We study odd-even mass staggering of nuclei, also called pairing gaps, using
a Skyrme self-consistent mean-field theory and a numerically exact treatment
of the pairing Hamiltonian. We find that the configuration-space Monte Carlo
method proposed by Cerf and Martin offers a practical computational
procedure to carry out the numerical solutions in large-dimensional model
spaces. Refitting the global strength of the pairing interaction for
443 neutron pairing gaps in our number-conserving treatment, we find the correction to the pairing correlation energies and pairing gaps to have rms values of 0.6 MeV and 0.12 MeV, respectively.  The exact treatment
provides a significant improvement in the fit to experimental gaps, although
it is partially masked by a larger rms error due to deficiencies in other
aspects of the theory such as the mean-field energy functional.
\end{abstract}


\maketitle

\section{Introduction}

Computer resources now make it possible to test theories of nuclear structure using
the entire body of nuclear data.  One particular aspect of nuclear structure
is pairing, which is important for determining stability and
dynamical properties of nuclei.  The Bardeen-Cooper-Schrieffer (BCS)
theory~\cite{BCS} has been a paradigm for treating nuclear pairing, but
it is not well justified in finite nuclei.  Besides its violation of
particle-number conservation, the condensate may collapse in finite systems.
 A recent global study of
nuclear pairing gaps~\cite{be09} found that $\sim$ 25\% of nuclei
lacked a BCS pairing condensate because of the weakness of the interaction
or a low single-particle density of states.  The observed smoothness of nuclear binding
energies calls for a theory that does not force a discontinuous jump between
ground states with and without pairing condensates.  In Ref.~\onlinecite{be09} it was found that a small but significant overall
improvement in theory could be
achieved by using the Lipkin-Nogami (LN) extension of BCS to correct
for particle-number violation~\cite{li60}. However, the LN
treatment has its own limitations. For example, it becomes inaccurate near
closed shells when implemented in the usual way \cite{do93,be00}.  On
a practical level, iterative BCS-LN solvers often have convergence problems near closed shells. We note that there are many methods other than the LN extension
of BCS to treat the pairing interaction more accurately, including
direct diagonalization in truncated spaces~\cite{pi05}.  For
methods that emphasize particle number conservation, see Ref.~\cite{st07}
and references therein.

Here we address the question of the importance of a better treatment of
pairing correlations by carrying out a global survey using a numerically
exact technique to calculate pairing correlation energies at fixed
particle number.  In particular, we employ the configuration-space
Monte Carlo (CSMC) algorithm of Cerf and Martin~\cite{ce93}. Numerically exact
solutions can also be obtained by direct diagonalization of the pairing
Hamiltonian in configuration spaces of fixed seniority ~\cite{ze05,mo97},
but the CSMC method is more efficient and can be implemented in much
larger spaces.

Since our aim is to assess the relative performance of the theory with and
without an exact treatment of pairing, we will avoid introducing extraneous
elements and closely follow the methodology of Ref.~\cite{be09}. In that
work the performance of various self-consistent mean-field (SCMF) methods
was tested on the neutron and proton pairing gaps for odd-$A$ nuclei; here
we use the same 443 odd-neutron gaps to assess the importance of an exact
treatment of the pairing interaction.

The neutron pairing gap for an (odd) neutron number $N$ is defined by the
second-order energy difference in neutron number
\be
\label{gap}
\Delta^{(3)}_o(N) = -\frac{1}{2} \left[ E(N+1)+E(N-1) - 2 E(N)\right] \;,
\ee
where $E(N)$ is the ground-state energy of the nucleus with $N$ neutrons and $Z$ protons. The proton number $Z$ is the same for all
three nuclei in Eq.~(\ref{gap}) and is not indicated explicitly in the formula.

As a prototype of an SCMF theory we use the energy density functional
constructed from the SLy4 Skyrme functional~\cite{ch98} for the normal
density part and a density-dependent contact interaction for the pairing
part. The Hartree-Fock+BCS (HF+BCS) equations are solved using the {\sc EV8}
code~\cite{bo05}. We construct a pairing Hamiltonian whose single-particle
orbital energies and pairing matrix elements are extracted from the SCMF
calculation of Ref.~\cite{be09}. Next, we solve this Hamiltonian exactly using the CSMC method,
which is free of a sign problem when all pairing matrix elements are
attractive.
The SCMF interaction energies are also taken
from the calculations of Ref.~\cite{be09}. The main difference here is in the
treatment of pairing correlation energies and the strength of the pairing
interaction.  The performance of the theory is measured by the
root-mean-square (rms) of the residuals with respect to the experimental data
set after making a least-squared fit of the overall pairing interaction
strength. The differences between the Monte Carlo treatment and the BCS
approximation are used to estimate the importance of a particle-number-conserving exact treatment of pairing.

The outline of this paper is as follows. In Sec.~\ref{methodology} we
discuss our methodology of constructing a pairing Hamiltonian from the SCMF
results and how we use its exact solution to obtain an improved estimate of
the pairing gap. In Sec.~\ref{CSMC} we describe the CSMC method used to
solve the pairing Hamiltonian. In Sec.~\ref{results} we present our results
for the pairing gaps. Our conclusions are given in
Sec.~\ref{conclusion}.

\section{Methodology}\label{methodology}

The leading approach in the search for a computationally tractable theory of
nuclear structure starts with an SCMF theory to
construct a set of configurations and then mixes the configurations through
a residual interaction to restore broken symmetries and add a correlation
contribution to the total energy.  By itself, the mean-field theory is
straightforward.  However, there are different ways to introduce
correlations, even if we limit ourselves to pairing correlations.  The BCS
treatment is the simplest way to introduce pairing correlations and can be
easily implemented once the single-particle wave functions and energies have
been obtained from the mean-field theory.  The Hartree-Fock-Bogoliubov (HFB)
approximation is an extension that is required when the orbital properties
depend on pairing, but like BCS it violates particle-number
conservation. To gain the benefit of the HFB approximation,
the pairing Hamiltonian must be
defined in very large model spaces and with a more general interaction than
can be treated with the present CSMC method.  Here we construct pairing
Hamiltonians in which the orbitals are fixed from the mean-field
calculation, as in the BCS approximation, and have interaction matrix elements that are
all attractive. The HFB might be required in the dripline region. However, since
very few of the known experimental gaps are in this region,
our conclusions should apply to
the vast majority of nuclei for which data exist.

For construction of the pairing Hamiltonian, we follow closely the treatment
of Ref.~\cite{bo85} as implemented in EV8.  The single-particle energies
are taken directly from the eigenvalues of the single-particle SCMF Hamiltonian and the interaction is chosen as a density-dependent contact interaction
\be\label{contact}
V(\mathbf{r},\mathbf{r'})=-V_0\left(1-\eta \frac{\rho(\mathbf{r})}{\rho_0}\right)\delta(\mathbf{r}-\mathbf{r'})
\ee
together with an energy cutoff factor described below.  In Eq.~(\ref{contact})
$\rho_0=0.16 \mbox{ fm}^{-3}$ is the conventional saturation density of nuclear
matter and the parameter $\eta$ controls the specific density dependence.
We will use $\eta=0.5$, called the
``mixed'' density-dependent pairing interaction. The strength $V_0$
is determined by minimizing the rms of
the residuals of the calculated pairing gaps from their experimental
counterparts~\cite{be09}.

As implemented in EV8, the mean field is invariant under time reversal
and the self-consistent single-particle orbitals appear in degenerate time-reversed
pairs $i$ and $\bar i$ with energy $\ei$. The total number of orbital pairs
is $\Omega$. The antisymmetrized
pairing matrix elements $V_{ij}$ are taken to be
\be
V_{ij} \equiv f_i (\langle i \bar i | V | j \bar j \rangle - \langle i \bar i | V | \bar j  j \rangle)  f_j \;,
\ee
where $V$ is given by Eq.~(\ref{contact}) and $f_i$ are energy cutoff factors~\cite{bo85}
\be
\label{cutoff}
f_i = \left[\frac{1}{1 + e^{(\varepsilon_i - a)/b}} \frac{1}{1 + e^{(-\varepsilon_i - a)/b}}\right]^{1/2}
\ee
with $a=5$ MeV and $b=0.5$ MeV.
Denoting the single-particle orbitals by $\phi_i({\bf r},\sigma)$, we have
\begin{eqnarray}
\label{Vij}
V_{ij} = -f_i f_j V_0 \int d{\bf r} \left( \sum_{\sigma}
| \phi_i ({\bf r}, \sigma)|^2 \right) & \left(
 \displaystyle \sum_{\sigma'} |\phi_j ({\bf r}, \sigma')|^2 \right ) \nonumber\\
& \times \left( 1-\eta \frac{\rho(\mathbf{r})}{\rho_0} \right)
\;.
\end{eqnarray}

Next we construct the pairing Hamiltonian
\be
\label{hamil}
\hat H = \hat H_1 + \hat H_2 = \sum_{i}^{\Omega} \ei (\adag_i a_i +\adag_{\bar{i}} a_{\bar{i}}) +
\sum_{i\neq j}^{\Omega} V_{ij} \adag_i \adag_{\bar{i}} a_{\bar{j}} a_{j} \;.
\ee
Note that the Hamiltonian (\ref{hamil})
does not include diagonal
matrix elements $V_{ii}$.  We assume that they have
already been incorporated into the mean-field part of the energy density functional.

An Hamiltonian of the form Eq.~(\ref{hamil}) was used in a recent study
comparing the BCS approximation with exact matrix diagonalization results in
model spaces of size $\Omega=16$~\cite{sa08}. However the sizes of the model
space required for a global survey are prohibitively large for direct matrix
diagonalization methods to be practical. We therefore use the CSMC method
which scales much more gently as a function of $\Omega$. This method can be
used to find the exact ground-state energy $E_{\rm CSMC}$ of
the Hamiltonian in Eq.~(\ref{hamil}) to within a statistical error. We note that the pairing Hamiltonian can be solved algebraically for special forms
of the interaction following Richardson's method~\cite{ri63}, but these are not
applicable to more general interactions such as in Eq.~(2).

Our improved estimate for the total ground-state energy is given by
\be
\label{etot}
E = E_{\rm SCMF} - E_{\rm BCS} + E_{\rm CSMC} \;,
\ee
where $E_{\rm SCMF}$ is the SCMF energy calculated with SLy4 plus the
density-dependent contact pairing interaction (\ref{contact}), and $E_{\rm BCS}$ is the BCS
ground-state energy of the Hamiltonian in Eq.~(\ref{hamil}). In
Eq.~(\ref{etot}) we are essentially replacing the BCS energy of
the Hamiltonian in Eq.~(\ref{hamil}) with its exact CSMC ground-state energy.  Both $ E_{\rm
SCMF}$ and $E_{\rm BCS}$ are calculated with an interaction strength $V_0$
determined by minimizing the rms deviation of the SCMF gaps from the
experimental gaps. However, $E_{\rm CSMC}$ is calculated with a renormalized
strength, determined by minimizing the rms residuals (with respect to
experiment) of the theoretical gaps calculated from Eq.~(\ref{etot}) and
Eq.~(\ref{gap}).

One problem of using a pairing Hamiltonian from a theory such as the one
discussed in Ref.~\cite{be09} is that there are diagonal interaction matrix
elements both in the mean-field part as well as in the pairing part of the
energy functional.  Since the pairing interaction is added to describe
correlations beyond those obtained in the SCMF with a single Slater
determinant, it should not add diagonal interactions beyond the mean field.
At an extreme, if the BCS condensate collapses, the BCS correlation energy
should be zero. Because of these considerations we do not include diagonal
interaction matrix elements $V_{ii}$ in our Hamiltonian (\ref{hamil}) for
either the CSMC or the BCS calculations. Such matrix elements remain, however, in the pairing part of the SCMF theory.

\section{Configuration space Monte Carlo solver}\label{CSMC}

The CSMC Hamiltonian solver has been applied to individual isotope chains~\cite{ca98}
but our work here is its first use in a global survey.  To introduce the
various parameters of the method and make our presentation self-contained, we
review the algorithm in Sec.~\ref{CSMC:method}. In Sec.~\ref{staterr}, we discuss the statistical Monte Carlo error and demonstrate the computational scaling
properties of the method.

\subsection{The Monte Carlo method}\label{CSMC:method}
In the following, we assume the particle number $N$ to be even. The
algorithm applies to Hamiltonians of the form in Eq.~(\ref{hamil}) for which
all pairing interaction matrix elements satisfy $V_{ij} \le 0$.  The overall
operation of the algorithm is similar to many other Monte Carlo methods
where a trial state $|\Phi\rb$ is evolved in imaginary time
\be
|\Psi (\tau)\rangle = e^{-(\hat H-E_t) \tau} |\Phi \rangle\;,
\ee
where $\hat H$ is the system's Hamiltonian and $E_t$ is an energy parameter
 adjusted to keep the normalization of $|\Psi(\tau)\rangle$
approximately fixed.  The initial evolution filters out the ground-state
component of the initial trial state, and subsequent evolution is used to
obtain better statistics for the ground-state energy.

The remaining details of the Monte Carlo method relies on the
representation of $\Psi(\tau)$ as a superposition of pure paired
configurations~\cite{ce93}. Let us label the fully paired eigenstates of $\hat H_1$, in Eq.~(\ref{hamil}), by $|{\bf
n}\rangle = |n_1,n_2,\ldots n_{\Omega}\rangle$ where $n_i=0$ or $1$ is the
pair occupation number of the $i$-th two-fold degenerate level, and
\be
\hat H_1 | {\bf n}\rb = E_{\rm sp}({\bf n}) |{\bf n}\rb
\ee
with
\be\label{Esp}
E_{\rm sp}({\bf n})= 2\sum_{i} \ei n_i \;.
\ee
The wave function $|\Psi(\tau)\rangle$ can be written as a linear
combination of these paired configurations $| {\bf n} \rangle$
\be
|\Psi(\tau)\rangle = \sum_{\bf n} \alpha_{\bf n}(\tau) |{\bf n}\rangle \;.
\ee
The coefficients $\alpha_{\bf n}(\tau)$ can be chosen to be all
positive and normalized as
\be
\sum_{\bf n} \alpha_{\bf n}(\tau) = 1 \;.
\ee
Thus, the wave function $|\Psi(\tau)\rangle$ can be represented by an
ensemble of paired configurations $| {\bf n}\rangle$ that are distributed
with probability $\alpha_{\bf n}(\tau)$.

The evolution in imaginary time is carried out as a series of time
evolutions, each of which is over a small time step $\Delta \tau$
\be
|\Psi(\tau + \Delta \tau) \rangle = e^{-(\hat H-E_t) \Delta \tau} |\Psi (\tau) \rangle \;.
\ee
Using the Suzuki-Trotter symmetric decomposition \cite{tr59, su77,
hi81}, we write the short-time propagator as
\begin{eqnarray}\label{st}
e^{-(\hat H-E_t) \Delta \tau} =
e^{-(\hat H_1 -E_t) \Delta \tau /2} e^{-\hat H_2  \Delta \tau}  e^{-(\hat H_1 -E_t) \Delta \tau /2}   \nonumber \\
+\mathcal{O}(\Delta \tau^3) \;.
\end{eqnarray}
We are interested in calculating the matrix elements of Eq.~(\ref{st})
between two paired configurations $| {\bf n}\rangle$ and $| {\bf n}'\rangle$.
Since $\hat H_1$ is diagonal in the
$|{\bf n}\rangle$ basis, the only non-trivial part is the matrix elements of
$e^{-\hat H_2\Delta \tau }$.
Expanding this propagator in a Taylor series, we
have~\cite{ce93}
\be
\label{tbme}
\langle {\bf n}' |e^{-\hat H_2 \Delta \tau} | {\bf n} \rangle
= e^\nu \displaystyle  \sum_{L=0}^{\infty}
P(L)
\left [\frac{1}{\omega^L}
\sum_{\kappa} W_{\kappa} ({\bf n} \to {\bf n}') \right ],
\ee
where $\omega=\frac{N}{2}(\Omega-\frac{N}{2}+1)$ and $\nu = \omega \bar{V}
\Delta \tau$ defines a dimensionless time step with $\bar{V}= \displaystyle
\sum_{ij} V_{ij}/\Omega^2$. Here $W_{\kappa} ({\bf n} \to {\bf n}')$
represents the weight of a path of $L$ pair hops that takes the
configuration ${\bf n}$ to ${\bf n}'$. Each pair hop describes the
transition of a pair of particles from an occupied two-fold level to an
empty two-fold level. The probability to have $L$ pair hops in the time
interval $\Delta \tau$ is a Poisson distribution $P(L) = e^{-\nu} \nu^L/L!$.
The parameter $\omega$ represents the total number of possible pair hops and
$\nu$ is the average number of pair hops in the time interval $\Delta \tau$.
The weight $W_{\kappa}({\bf n} \to {\bf n}')$ is given by
\be\label{Wkappa}
W_{\kappa}({\bf n} \to {\bf n}') = \displaystyle \prod_{m=1}^{L} \left |\frac{ V_{i_m j_m}}{\bar{V}} \right | \;,
\ee
where $(i_m, \bar{i}_m)$ and $(j_m, \bar{j}_m)$ are the orbital pairs whose
occupations are swapped at the $m$-th step of the $L$-step pair hop process.

In practice, we carry out the Monte Carlo evolution as follows. We take the
initial state $|\Phi\rb$ to be the ground-state configuration of $\hat H_1$ and
replicate it $N_e$ times to generate the initial ensemble.  Subsequently,
the members of the ensemble are evolved independently. For each time step
$\Delta \tau$, the time evolution is done stochastically using
Eqs.~(\ref{tbme}) and (\ref{Wkappa}).
The number of pair hops $L$ is drawn from the Poisson distribution $P(L)$,
and an $L$-step pair hop process is carried out. At each step we choose an
occupied pair orbital $(i_m, \bar{i}_m)$ from a uniform distribution,
and an orbital $(j_m, \bar{j}_m)$ which is either unoccupied or equal to
$(i_m, \bar{i}_m)$, again from a uniform distribution. The
occupation numbers of $(i_m, \bar{i}_m)$ and $(j_m, \bar{j}_m)$ are swapped.
The resulting new configuration is replicated stochastically with a weight
of $\exp[-(E_{\rm sp}({\bf n}) + E_{\rm sp}({\bf n}') -2 E_t) \Delta \tau /2]
W_\kappa({\bf n} \to {\bf n}')$.

We adjust the normalization energy $E_t$ during the time evolution to keep
the ensemble size stable. At the $k$-th time step we define
\be
E_t(k)= E_t(k-1)+\frac{1}{\Delta \tau} \ln \left [\frac{N_e(k-1)}{N_e(k)}\right]\;,
\ee
where $N_e(k)$ is the size of the ensemble at the $k$-th time step.

We repeat the above process $N_{\tau}$ times for the initial evolution.
The resulting ensemble of fully paired

configurations ${\bf n}_m$ ($m=1,\ldots,N_e$) can be expressed as the wave
function
\be\label{Psi1}
|\Psi_1 \rangle = \frac{1}{N_e}\sum_{m}^{N_e} |{\bf n_m}\rangle
\ee
and is our first representative of the ground-state ensemble.

The ground-state energy $E$ can be calculated from the ground state $\Psi$ using
$E = \sum_{{\bf n}'} \langle {\bf n}' |H | \Psi \rangle$ (where we have used
$\sum_{{\bf n}'} \langle {\bf n}' | \Psi \rangle=1$). Approximating  $\Psi$ by  $\Psi_1$ in Eq.~(\ref{Psi1}), we estimate the ground-state energy to be
\be\label{E1}
  E_1  = \frac{1}{N_e}\sum_{m=1}^{N_e}
\left[ E_{\rm sp}({\bf n}_m)+ E_v ({\bf n}_m)\right] \;,
\ee
where $ E_{\rm sp}({\bf n})$ is given by Eq.~(\ref{Esp}) and
\be\label{Ev}
E_v ({\bf n})= \sum_{{\bf n}'}\langle {\bf n}' | H_2 | {\bf n} \rangle \\
         = \displaystyle \sum_{ij}' V_{ij} \;.
\ee
The prime on the summation in Eq.~(\ref{Ev}) denotes  that the sum is
restricted to those combinations $ij$ where the orbital pair $(i,\bar{i})$
is occupied in $| {\bf n} \rangle$ and the orbital pair $(j,\bar{j})$
is either unoccupied in $| {\bf n} \rangle$ or the same as $(i,\bar{i})$.

Additional representatives $|\Psi_i \rangle$ of the ground-state wave
function are generated by evolving the ensemble an additional number of time
steps $N_T$ and taking a representative every $N_c$ steps to ensure
uncorrelated ensembles. Using relations similar to Eq.~(\ref{E1}), we obtain
$N_E=N_T/N_c$ estimators $E_1, E_2, \ldots, E_{N_E}$ for the ground-state
energy. The final estimate for the CSMC ground-state energy is
\be\label{emc}
E_{\rm CSMC} = \frac{1}{N_E} \sum_i^{N_E} E_i  \;,
\ee
and its corresponding statistical error is
\be
\sigma = \sqrt {\frac{1}{N_E (N_E-1)} \sum_i^{N_E} (E_i - E_{\rm CSMC})^2 }\;. \label{sigma}
\ee

So far, we have discussed a system with an even number of particles $N$.
The generalization to an odd number $N_o$ is straightforward. We put a
single particle in one of the orbitals of a degenerate pair.
This pair of orbitals becomes effectively blocked, i.e.,
 it cannot participate in the pair transitions between orbitals.
The energy of the remaining $N_o-1$ particles
is found by applying CSMC to the reduced space in which the blocked orbital
pair is excluded. The total energy of the $N_o$-particle system is then given
by
\be
\label{bln}
E _{b}(N_o) = E_{b} (N_o-1) + \varepsilon_{b} \;,
\ee
where $\varepsilon_{b}$ is the single-particle energy of the blocked orbital and $E_{b}
(N_o-1)$ is the energy of $N_o-1$ particle system in the reduced space.  The
ground-state energy of the odd-$N$ system is found by minimizing
Eq.~(\ref{bln}) over different choices of the blocked orbital $b$.

For the calculations in this work, we have taken  $N_\tau=5000$,
$N_T=50,000$, $N_c=500$ and $N_e(0)=25,000$.  This gives $N_E=100$ estimators
$E_i$ of the the ground-state energy and its error in Eqs.~(\ref{emc}) and
(\ref{sigma}), respectively.

\subsection {Statistical error}\label{staterr}

The statistical error $\sigma$ in the CSMC energy estimate can be written as
\begin{equation}\label{MC-err}
\sigma=\sqrt{\frac{\chi_e}{N_E N_e}}\;\,\sigma_{\rm in} \;,
\end{equation}
where $\sigma_{\rm in}^2$ is the intrinsic variance of the energy, i.e., the variance of the quantity $E_{\rm sp}({\bf n})+E_v({\bf n})$ for paired configurations ${\bf n}$ that are distributed according to $\alpha_{\bf n}$.
Here $N_E=N_T/N_c$ is the number of uncorrelated ensembles of size $N_e$ used in the CSMC calculation.
The replication process described in Sec.~\ref{CSMC:method} introduces correlations between configurations in the ensemble at a given time step and $N_e/\chi_e$ (with $\chi_e >1$) represents the effective number of uncorrelated configurations.

 In the following, we provide an estimate for the intrinsic standard deviation $\sigma_{\rm in}$ assuming a constant pairing interaction. In this case $E_v({\bf n})$ in Eq.~(\ref{Ev}) is a constant, and $\sigma^2_{\rm in}$ is determined solely by the variance of the  $E_{\rm sp}({\bf n})$.
With $E_{\rm sp}({\bf n})$ given by Eq.~(\ref{Esp}), its average value $\bar E_{\rm sp}$ over the various configurations in the ensemble is
\begin{align}
\bar E_{\rm sp}= 2\sum_{\bf  n} \sum_{i=1}^{\Omega} \ei  n_i \alpha_{\bf n}
           = \lambda N + 2\sum_{i=1}^{\Omega} (\ei - \lambda) \bar{n}_i \;,\label{sesp}
\end{align}
where $\bar{n}_i = \displaystyle \sum_{\bf n} n_i \alpha_{\bf n}$.\footnote{Note that ${\bar n}_i$ differs from the quantum mechanical expectation value of the pair occupation operator $\hat n_i$.} Eq.~(\ref{sesp}) holds for any constant $\lambda$ but we choose $\lambda$ to be the chemical potential to minimize the particle number fluctuations.

In the Appendix we use the BCS wave function to estimate the fluctuations in $E_{\rm sp}$ [see Eq.~(\ref{insbcs})]. For a uniform single-particle spectrum with a bandwidth $E_c \gg \Delta$ ($\Delta$ is the BCS pairing gap), we find
\begin{equation}
\sigma_{\rm in}^2 \approx \frac {1}{2} \Omega E_c \Delta \;.  \label{insdel}
\end{equation}
We can use this expression to estimate the scaling of $\sigma_{\rm in}$ with the size $\Omega$ of the single-particle space. For weak to moderate pairing, $\Delta \propto \Omega$, and $\sigma_{\rm in} \propto \Omega^{3/2}$. Our simple estimate (\ref{insdel}) is accurate to within a factor of $\sim 2$ (see Fig.~\ref{inserr} in the Appendix).
We have checked that even in cases when the single-particle spectrum is non-uniform and the pairing interaction is orbital-dependent (e.g., the nuclear pairing Hamiltonians used in this work), expression (\ref{insdel}) (where $\Delta$ is taken to be an average pairing gap) provides a reasonable estimate for the intrinsic error in $E_{\rm sp}$.

\begin{figure}[tb]
\begin{center}
\includegraphics[width=\columnwidth]{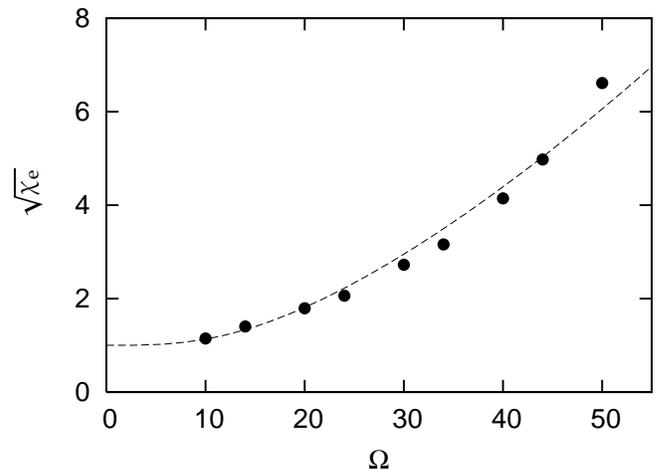}
\caption{The factor $\sqrt{\chi_e}$ as a function of $\Omega$ for an equally-spaced single-particle spectrum with level spacing of $1$ MeV and pairing strength of $V_{ij}=0.3$ MeV. The dashed line describes the fit $\chi_e = 1+(\Omega/\Omega_0)^3$ with $\Omega_0=15.2$.}
\label{sqrt_chi}
\end{center}
\end{figure}

If all $N_e$ configurations of the ensemble at a given time step were to be uncorrelated, the CSMC error would have been  $\sigma_{\rm in}/\sqrt{N_E N_e}$. Since these configurations are correlated in the CSMC calculation, the actual statistical  error is larger by a factor of $\sqrt{\chi_e}$ [see Eq.~(\ref{MC-err})]. In Fig.~\ref{sqrt_chi} we show (solid circles) this enhancement factor $\sqrt{\chi_e}$ (as determined empirically from the CSMC statistical error for $\nu=0.1$) versus $\Omega$ for a uniform single-particle spectrum with level spacing of $1$ MeV at half filling ($N=\Omega$) and a constant pairing strength of $V_{ij}=0.3$ MeV.  The dashed line is a fit to $\chi_e \sim 1+(\Omega/\Omega_0)^3$. In general we find that the scaling of $\chi_e$ with $\Omega$ depends on the strength of the pairing interaction.

\begin{figure}[htb]
\begin{center}
\includegraphics[trim= 0   0.75  0  15, clip=true, width=\columnwidth]{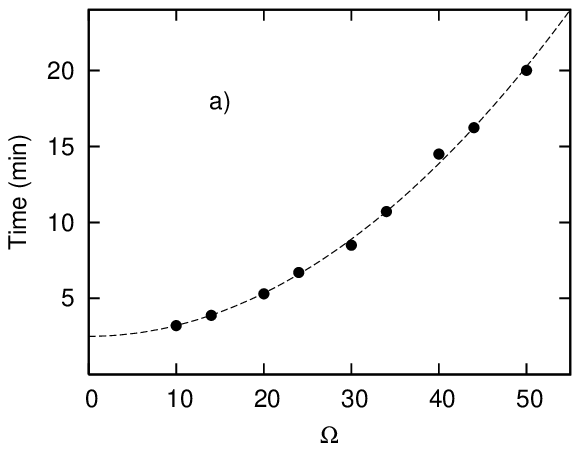}
\includegraphics[trim= 0   0  0  0, clip=true, width=\columnwidth]{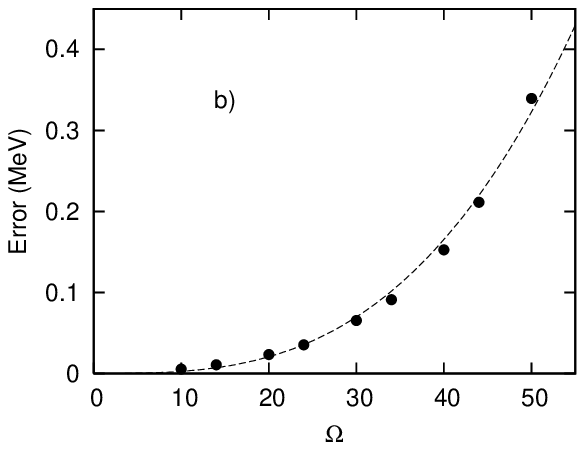}
\caption{Scaling of computational effort with the number of
orbitals $\Omega$ for an equally-spaced single-particle spectrum with level
spacing of 1 MeV and a constant pairing interaction $V_{ij}=0.3$ MeV: a)  single-processor CPU time for a single CSMC calculation (the dashed line is a fit describing a scaling of $\Omega^2$); b) statistical error of Eq.~(\ref{sigma}) (the dashed line corresponds to a scaling of $\Omega^3$). See text for the values of the CSMC parameters.}
\label{tvo}
\end{center}
\end{figure}

To illustrate the scaling of the CSMC computational time with the size
$\Omega$ of the single-particle space, we consider the same example as in Fig.~\ref{sqrt_chi}.
The results, shown by symbols in the upper panel of
Fig.~\ref{tvo}, scale (up to an additive constant) as $\Omega^2$
(dashed curve).

The lower panel of Fig.~\ref{tvo} shows the statistical
error calculated from Eq.~(\ref{sigma}).  It appears to scale as $\Omega^3$ (dashed
line)\footnote{ This error seems to depend on the parameters of the pairing
Hamiltonian and for a stronger pairing interaction we find a more moderate
scaling of $\Omega^2$.}.  Spaces as large as $\Omega=30$ are easily computed
and we shall argue below that the accuracy achieved is adequate for our purposes.
In contrast, if the calculations were done by conventional matrix diagonalization, one would have to deal with a matrix of dimension
$1.6 \times 10^8$ and $\Omega=50$ (corresponding to a matrix of dimension
 $1.3 \times 10^{14}$) would be completely out of reach.

The CSMC calculations for the global survey were carried out using two values of $\nu$
(0.05 and 0.10) and averaging their respective energy estimates. We checked that the corresponding time steps were sufficiently small to
avoid a significant systematic error from the Suzuki-Trotter decomposition
in Eq.~(\ref{st}).
To test for other biases in the CSMC algorithm,
we compared with the matrix
diagonalization of the Hamiltonian (\ref{hamil}) for two cases presented in
Ref.~\cite{sa08}, namely $^{118}$Sn and $^{206}$Pb.  The pairing
Hamiltonians were obtained from Ref.~\cite{be09}; $\hat H_1$ is derived from SCMF
with the Skyrme SLy4 energy functional and $\hat H_2$ is of the contact form as
in Eq.~(\ref{contact}) with $\eta=0$ (and no cutoff factors $f_i$). The
single-particle space in these examples has a size $\Omega=16$, which
requires matrices of dimension $\sim 13,000$ for the direct diagonalization.
The calculated correlation energies (measured relative to
the HF ground-state energy) are shown in Table~\ref{exact} for two values of
the pairing strength, $V_0=$ 360 and 450 MeV fm$^3$. From the results we see
that there are no discernible systematic errors in the CSMC calculations.
\begin{table}[htb]
\begin{center}
\begin{tabular}{|cc|cc|cc|}
\hline
                       &    &        \multicolumn{2}{c|}{$V_0=360$}     & \multicolumn{2}{c|}
{$V_0=450$} \\
                       &                      &           Lanczos          &     CSMC          &  Lanczos    &     CSMC       \\
\hline
$^{118}$Sn &            &    $2.564$       & $2.569 \pm 0.006$   &  $4.553$ &  $4.546 \pm 0.006$ \\
$^{206}$Pb &            &    $0.363$       & $0.365 \pm 0.004$   &  $0.626$ &  $0.626 \pm 0.005$ \\
\hline
\end{tabular}
\caption{Comparison of pairing correlation energies calculated by CSMC and by exact
diagonalization method (Lanczos algorithm). The interaction strength
$V_0$ is in units of MeV-fm$^3$ and the correlation energies in units of MeV.}
\label{exact}
\end{center}
\end{table}

For the CSMC calculation in our global survey, it is important that the
Monte Carlo statistical error does not degrade the accuracy of the
calculated pairing gaps to a point where the performance measure would be
affected.  The maximal permissible statistical error is estimated as follows.
 We take a typical rms of the residuals (deviations between theory and experiment)
 in the range 0.25-0.30 MeV,
and demand that the Monte Carlo statistical contribution, calculated in
quadratures, be less than 0.01 MeV.  This requires that the average
statistical error for the pairing gap be smaller than
$\sqrt{0.25^2 -0.24^2} \approx 0.07$ MeV.  In fact, with our choices of the
numerical parameters in the global calculation, the maximal statistical
error ($\sim 0.05$ MeV) satisfies this upper bound for all cases.

\section{Results}\label{results}

For our global survey of odd neutron gaps, we take the same nuclei as in
Ref.~\cite{be09}, where 443 odd neutron pairing gaps were calculated and
compared with experiment. Our procedure for obtaining a new set of
theoretical gaps involves the following steps:

\begin{enumerate}
  \item We start with the full SLy4+pairing energies as calculated in
  Ref.~\cite{be09} and construct the pairing Hamiltonian $\hat H$ in
  Eq.~(\ref{hamil}) using the converged SCMF single-particle energies and
  wave functions. The diagonal interaction matrix elements $V_{ii}$ are not included in Eq.~(\ref{hamil}).
  \item  We calculate the BCS ground-state energy of $\hat H$, taking the same
    interaction strength $V_0$ as in the original calculations (i.e.,
    $V_0 = 700$ MeV-fm$^3$),
    and subtract it off the total SCMF energy.
  \item We calculate the exact ground-state energy of $\hat H$ by CSMC and add it
   back to to obtain our new estimate of the ground-state energy. The overall
   interaction strength is renormalized in the CSMC calculation, and its value is
    determined by minimizing the rms residuals of the newly calculated pairing gaps.
  \item  To make a fair comparison with the BCS, we repeat step 3 with BCS
    energy (excluding diagonal interaction matrix elements as in the CSMC),
    refitting the overall strength of the interaction to minimize the rms
    residuals.\footnote{In principle, step 4 should not be necessary but we found
     that the strength
    $V_0$ reported in Ref.~\cite{be09} is not optimal for the BCS theory
    presented there.}
\end{enumerate}

Some remarks are in order regarding our refit. The CSMC calculations
were performed at two different values of $V_0$ (560 and 700 MeV fm$^3$). We
used a linear interpolation to obtain the ground-state energy for
interaction strengths between these two values.  The new value of $V_0$ is
determined by minimizing the rms of the residuals using a linear
least-squared fit.
The model space for our CSMC calculation consists of all orbitals for which
$f_i^2 > 0.01$ [see Eq.~(\ref{cutoff})]. We verified the convergence of our calculations by repeating
them for a model space with $f_i^2>0.001$. The largest model space used in
these calculations is $\Omega=64$ for the nucleus $N=156$ and $Z=100$. For
this nucleus each CSMC calculation takes about 30 minutes on a single
processor.

The results for our refits are shown in Table~{\ref{rmscom}} along with
the fit reported in Ref.~\cite{be09}.  First we note that the fitted
value of the interaction strength $V_0$ is smaller for the CSMC gaps than its fitted value for the BCS gaps
by about 6\%.  It is not surprising that the required strength is higher in
a theory (e.g., BCS) that is subject to pairing collapse and gives a zero correlation
energy in some of the nuclei.  In fact, the differences between correlation
energies comparing the CSMC and BCS can be quite large; their rms difference
is $\sim 0.6 $ MeV for the more than 900 nuclei in our data set.
However, the observable quantities are not the
correlation energies but the pairing gaps, for which the differences
 are much smaller.  The rms of the differences between
the CSMC and the BCS pairing gaps is 0.12 MeV.  Given that the total
rms residuals of the theory with respect to experiment is of the order
0.25-0.30 MeV, this difference between CSMC and BCS appears to be quite
significant.  However, one must realize that when there are independent
sources of error, the larger ones can effectively mask the others.  This
can be seen in the third column of table II, reporting the rms residuals
of the CSMC and BCS pairing gap with respect to experiment.  The corresponding values, 0.28 and
0.24 MeV, only differ by 0.04 MeV.  However, this is close
to what one would expect when adding in quadratures the error of the BCS
approximation (0.12 MeV) and the other sources of error (0.24 MeV).
We also note that the values for the fitted strength and the rms residuals
reported in Ref.~\cite{be09} are somewhat higher than their corresponding
values in our BCS fit, to which it should be compared.
\begin{table}[htb]
\begin{center}
\begin{tabular}{|c|c|c|}
\hline
$\quad$ Method $\quad$         &  $\quad V_0 \quad $  &  $\quad$  rms $\quad$   \\
                               &   (MeV fm$^3$)         &    (MeV)     \\
\hline
SCMF  \cite{be09}           &  700    &    0.30  \\
BCS             &  667    &    0.28  \\
CSMC            &  627    &    0.24  \\
\hline
\end{tabular}
\caption{The rms residuals of the calculated pairing gap using different
theoretical methods. See text for a
description of the various methods.}
\label{rmscom}
\end{center}
\end{table}

\begin{figure}[tb]
\begin{center}
\includegraphics[width=\columnwidth]{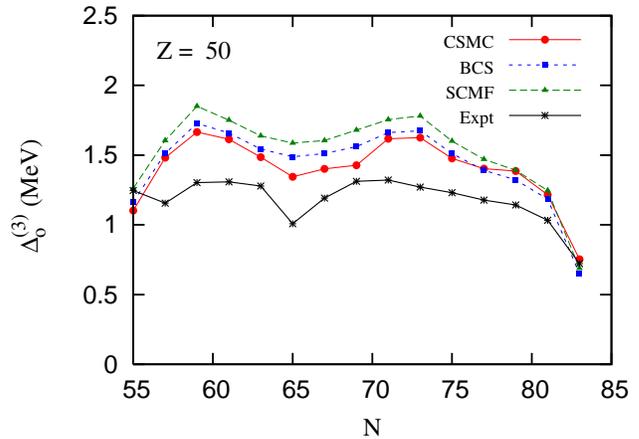}
\caption{Pairing gaps of Sn isotopes: predictions of various theories
(SCMF, BCS and CSMC) are compared with the experimental gaps.}
\label{sn}
\end{center}
\end{figure}

To illustrate the performance of the theory locally, we compare in
Fig.~\ref{sn} the theoretical and experimental pairing gaps for the chain of
Sn isotopes ($Z=50$).  All three theories (SCMF, BCS and CSMC) overestimate
the gap but follow correctly its overall dependence on neutron number,
including the the dip at
$N=65$ and the sharp drop near the $N=82$ shell closure. When compared with
 the experimental gaps, the CSMC shows a modest
but systematic improvement over the SCMF and BCS theories, except for
 the $N=77-81$ nuclei in the vicinity of the shell closure.

\begin{figure}[tb]
\begin{center}
\includegraphics[width=\columnwidth]{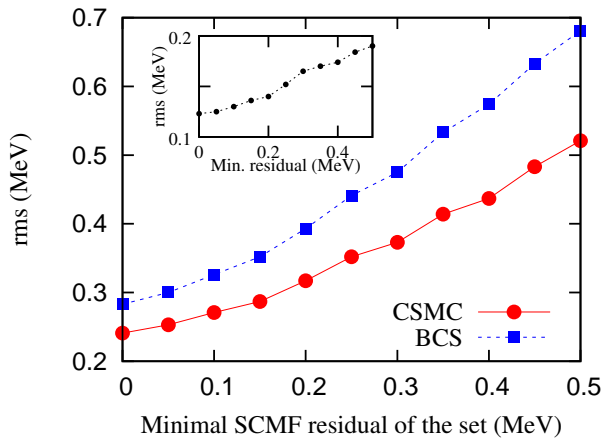}
\caption{The rms residuals of the pairing gap in nuclei
for which the absolute deviation of the SCMF pairing gap from the experimental
value is greater than the value shown on the horizontal axis. The inset shows
the rms of the CSMC correction to the BCS gap, the horizontal axis being the same as
the main graph.}
\label{greater}
\end{center}
\end{figure}

It would be useful to know whether there are any systematic criteria for
identifying nuclei for which the improved treatment of pairing has the
most benefit.  One criterion could be the magnitude of the error (i.e., residual)
 comparing the SCMF or BCS pairing gaps with their experimental values.
 To examine the dependence on the SCMF
error we take subsets of gaps whose SCMF residual is larger (in absolute value)
than some given value and calculate the rms of the subset as a function of the
lower cutoff.  The results are shown in Fig.~\ref{greater}.
The rms error of the subset increases with cutoff but in the CSMC approach
it does so at a lower rate than in the BCS treatment. For example, when we keep
only those nuclei whose SCMF rms residual is greater than 0.5 MeV, we find that
the BCS rms increases to 0.68 MeV
while the CSMC rms is only 0.52 MeV, an improvement of 0.14 MeV.
The inset of Fig.~\ref{greater} shows the rms of the CSMC
correction to the BCS residual versus the lower cutoff of the SCMF residual.
This rms exhibits a gradual increase from about 0.12 MeV when all the nuclei are included to
about 0.19 MeV when we include only those nuclei whose SCMF residual is greater than 0.5
MeV.  Thus, there is a mild increase in the benefit derived from the exact
treatment when the residual error is large.

\begin{figure}[tb]
\begin{center}
\includegraphics[width=\columnwidth]{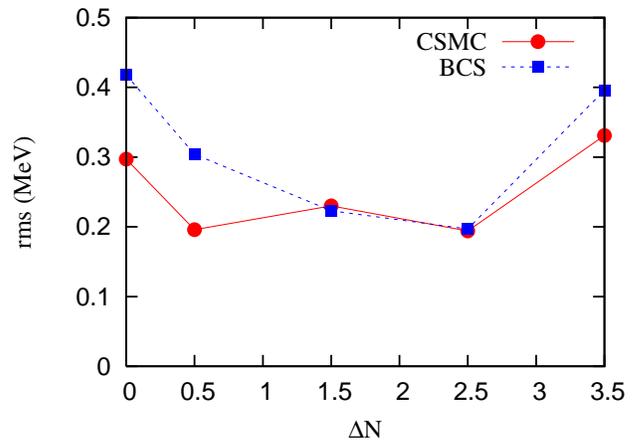}
\caption{ The rms residuals of the pairing gap as a
function of particle-number fluctuation $\Delta N$
[see Eq. (\ref{fluct})]. The nuclei were divided into bins of width 1 according
 to the values of $\Delta N$ obtained from the $v$ amplitudes of the SCMF theory
 ($\Delta N=0$ nuclei have their own bin).
 The points are positioned at the center of the bins and the lines are drawn to guide the eye.
}
\label{dn}
\end{center}
\end{figure}

To narrow further the conditions under which an exact treatment is beneficial,
we go back to the symmetries that are broken in mean field and BCS theory,
namely particle-number
conservation and rotational symmetry.  A measure of the BCS violation of particle-number
conservation is given by
\be\label{fluct}
(\Delta N)^2 \equiv \langle(\hat N - \langle \hat N \rangle)^2 = 4 \sum_{i>0} (1-v^2_i) v_i^2 \;,
\ee
where $v_i^2$ are the BCS occupation numbers. We divide the nuclei with odd
number of neutrons into bins of width 1 according to their particle-number
fluctuation $\Delta N$ (nuclei with $\Delta N=0$ have their own).
 Fig.~\ref{dn} shows the
rms of the residuals for the nuclei in each bin versus the midpoint of the bin.
The bin with $\Delta N=0$ consists of all nuclei for which the BCS
pairing has collapsed. Clearly the CSMC treatment is needed in that situation. The CSMC also gives an improvement for the bin centered at $ \Delta N = 3.5$ having the strongest pairing condensate.  This is likely due to the too-large pairing strength
 $V_0$ required for the global BCS fit.  Thus, when compared to the
exact CSMC results, the BCS approximation seems to be adequate when $1 \leq \Delta
N \leq 3$. If we use in the BCS treatment the lower value of the CSMC interaction strength, we find the BCS performance to improve gradually with increasing $\Delta N$ and to  become comparable to the CSMC performance for $\Delta N \geq 2$.

\begin{figure}[tb]
\begin{center}
\includegraphics[width=\columnwidth]{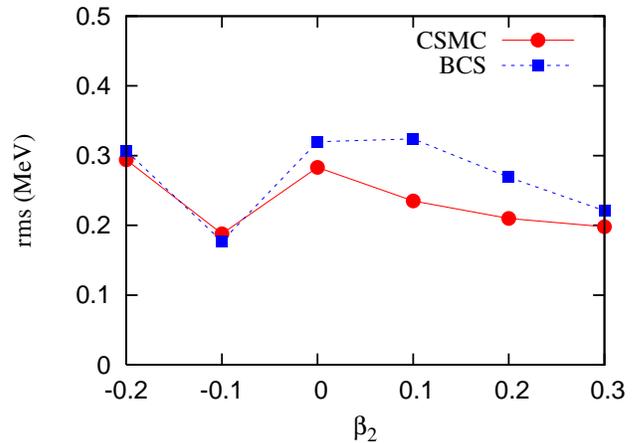}
\caption{ The rms residuals of the pairing gap as a function of deformation
$\beta_2$ [see Eq.~(\ref{beta2})]. The nuclei were divided into bins according
to their $\beta_2$ value in the SCMF treatment.  The points are positioned
at the center of the bins, and the lines are drawn to guide the eye.}
\label{def}
\end{center}
\end{figure}

The violation of rotational symmetry in a nucleus is often characterized by
the mass quadrupole deformation parameter $\beta_2$ defined by
\be
\label{beta2}
\beta_2=\sqrt{\frac{\pi}{5}} \frac{Q_0}{A \langle r^2\rangle} \;,
\ee
where $A$, $\langle r^2\rangle$ and $Q_0$ are, respectively, the mass
number, rms radius and intrinsic quadrupole moment of the nucleus. We divide
the odd-$N$ nuclei into bins of width 0.1 according to their deformation
$\beta_2$ in the SCMF treatment. The rms of the residuals are calculated for
the nuclei in each bin and their values (in both BCS and CSMC) are plotted
versus the bin centers in Fig.~\ref{def}. We observe that there is almost no
difference between the two treatments for oblate nuclei. For
spherical and strongly deformed prolate nuclei, the CSMC gives a moderate
improvement over BCS while for moderately deformed prolate nuclei there is a
significant improvement in the CSMC method as compared with the BCS
approximation.

\begin{figure}[tb]
\begin{center}
\includegraphics[width=\columnwidth]{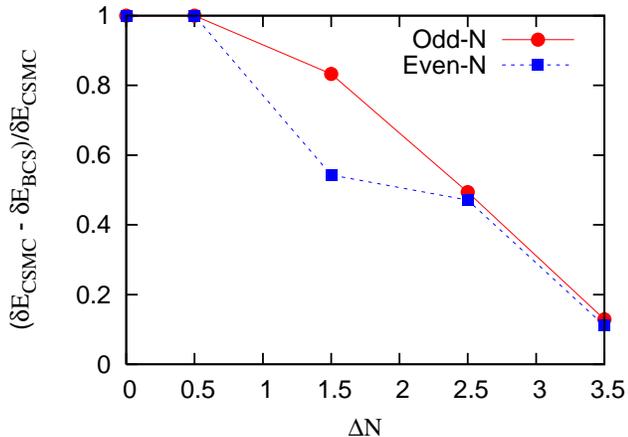}
\caption{ The ratio $(\delta E_{\rm CSMC} - \delta E_{\rm BCS})/\delta E_{\rm CSMC}$ as a function of particle-number fluctuation $\Delta N$ for nuclei with odd and even number of neutrons $N$. This ratio decreases as the BCS approximation becomes better at larger $\Delta N$. The case $\Delta N=1.5$ is a borderline case: the odd-$N$ behaves more like $\Delta N =0$ while the even-$N$ is closer to $\Delta N \gg 1$.
}
\label{corr}
\end{center}
\end{figure}

In Fig.~\ref{corr} we show the average value of the ratio between the CSMC correction to the BCS correlation energy and the CSMC correlation energy, $(\delta E_{\rm CSMC} - \delta E_{\rm BCS})/\delta E_{\rm CSMC}$, versus $\Delta N$ (here $\delta E_{\rm BCS} \equiv E_{\rm HF}- E_{\rm BCS}$ and $\delta E_{\rm CSMC} \equiv E_{\rm HF} - E_{\rm CSMC}$).  At $\Delta N=0$ the BCS solution collapses to the HF solution and this ratio is just $1$. We observe the above ratio to decrease monotonically versus $\Delta N$ as the BCS approximation becomes better.

\section{Conclusion}\label{conclusion}

Starting from an SCMF theory of the pairing gaps and treating pairing
correlations exactly beyond the BCS approximation (with a renormalized
pairing interaction strength), we found a significant improvement in the
 theory as measured by the rms residuals of the pairing gaps.
The exact calculations were carried out by constructing a pairing Hamiltonian
from the SCMF output and using the configuration space Monte Carlo (CSMC) method.

We find the improvement in the rms residuals of the pairing gaps to be most
significant in nuclei for which the BCS
condensate was weak, as measured by the smallness of particle-number
fluctuation $\Delta N$.  Based on our results, the BCS seems to be adequate
if one limits the theory to nuclei for which $1 \leq\Delta N \leq 3$. The artificially high value of the BCS interaction strength in a global fit leads to pairing gaps that are on average too large in nuclei with $\Delta N \geq 3$. We also found
the improvement to be larger in moderately deformed ($\beta_2 \sim
0.1-0.2$) prolate nuclei.

The total residual in the SCMF treatment of pairing gaps can be thought of
as coming from two parts, the inadequacy in the mean field and the
approximate treatment of pairing. Since in the CSMC method the pairing part
is treated exactly, the CSMC residuals are wholly due to the inadequacy of
the mean field. The rms of the residuals of the pairing gaps in the BCS
treatment over their values in the CSMC treatment is then an estimate of
the error involved in the BCS approximation. We find this rms value to be
$\sim 0.12$ MeV (see inset of Fig.~\ref{greater}), and propose it as the
bound on the accuracy that can be achieved in an SCMF theory that treats
pairing correlations approximately.

From the computational perspective, the most notable aspect of this work
is the use of the CSMC algorithm, which has not been used previously in a global survey.

\section*{Acknowledgements}
 We would like to thank R.~Capote for
providing us with the initial version of the Monte Carlo program, and T. Duguet for careful reading of the manuscript. This work
was supported in part by the U.S. Department of Energy under Grants
DE-FG-0291-ER-40608 and DE-FG02-00ER41132, and by the National Science
Foundation under Grant PHY-0835543. Computational cycles were provided by
the Bulldog clusters of the High Performance Computing facility at Yale
University and the Athena cluster at the University of Washington.

\appendix*
\section{}

In this Appendix we derive an estimate of the intrinsic statistical error $\sigma_{\rm in}$ based on the BCS wave function and show that it reduces to Eq.~(\ref{insdel}) in the limit of a large bandwidth $E_c \gg \Delta$ (assuming a uniform single-particle spectrum).

\begin{figure}[tb]
\begin{center}
\includegraphics[width=\columnwidth]{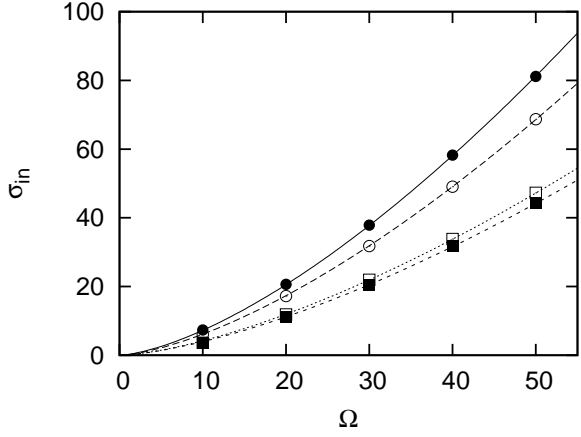}
\caption{The intrinsic error $\sigma_{\rm in}$ as a function of $\Omega$  for a uniform single-particle spectrum with level
spacing of 1 MeV and a constant pairing interaction $V_{ij}=0.3$ MeV. The exact intrinsic error (solid circles) calculated directly from CSMC (using a time step of $\nu=0.1$) is compared with the estimate of Eq.~(\ref{sigin2}) with the exact $\sigma^2_{n_i}$ calculated in CSMC (open circles).  The solid squares describe the BCS estimate Eq.~(\ref{insbcs}) while the open squares correspond to Eq.~(\ref{insdel}). The various lines are fits describing a scaling of $\Omega^{3/2}$}.
\label{inserr}
\end{center}
\end{figure}

Using Eq.~(\ref{sesp}) and assuming the pair occupation $n_i$ to be uncorrelated, we have
\begin{equation}
\label{sigin2}
\sigma^2_{\rm in} =  4\sum_i (\varepsilon_i-\lambda)^2 \sigma^2_{n_i} \;.
\end{equation}
where $\sigma^2_{n_i}$ is the variance of the pair occupation $n_i$.
To demonstrate the validity of Eq.~(\ref{sigin2}), we compare in Fig.~\ref{inserr} the exact intrinsic variance $\sigma^2_{\rm in}$ (solid circles)  with the r.h.s. of  Eq.~(\ref{sigin2}) (open circles). These quantities can be calculated directly in CSMC.
The results shown are for a uniform single-particle spectrum with a level spacing of 1 MeV at half filling and a constant pairing interaction of $V_{ij}=0.3$ MeV using a time step of $\nu=0.1$.  We have checked that Eq.~(\ref{sigin2}) remains a good approximation for the intrinsic error in more general cases, e.g., away from half filling and for a non-uniform spectrum.

Since $n_i$ can take only two values $0$ or $1$, the variance $\sigma^2_{n_i}$ of a given pair occupation is completely determined by its average
\begin{equation}\label{nivar}
\sigma^2_{n_i} = \bar{n}_i (1-\bar{n}_i ) \;.
\end{equation}

Next, we use the BCS wave function to estimate
\begin{equation}
\label{nibar}
\bar{n}_i=\frac{v_i}{u_i+v_i} \;,
\end{equation}
where $u_i$ and $v_i$ are the usual BCS amplitudes. Combining Eqs.~(\ref{sigin2}), (\ref{nivar}) and (\ref{nibar}), we have
\begin{align}
\sigma^2_{\rm in} &= 4 \sum_i (\varepsilon_i-\lambda)^2 \frac{u_i v_i}{(u_i+v_i)^2}  \nonumber \\
&= 2 \sum_i \Delta_i (\varepsilon_i-\lambda) \frac{u_i - v_i}{u_i+v_i} \;,
\label{insbcs}
\end{align}
where $\Delta_i$ are the level-dependent BCS pairing gaps.  The BCS estimate (\ref{insbcs}) for $\sigma_{\rm in}$ is shown in Fig.~\ref{inserr} by solid squares.

The BCS expression (\ref{insbcs}) for the intrinsic variance can be further simplified for a uniform single-particle spectrum (in which case the gap $\Delta$ is level-independent) with a bandwidth $E_c \gg \Delta$. In this limit most single-particle levels satisfy $|\epsilon_i - \lambda | \gg \Delta$ and the BCS amplitudes can be replaced by their non-interacting values. This leads to the simple expression in Eq.~(\ref{insdel}).

In Fig.~\ref{inserr} we also compare the BCS estimate (\ref{insbcs}) (solid squares) with its simplified version (\ref{insdel}) (open squares). The latter slightly overestimates the BCS expression; both underestimate the exact intrinsic variance but provide a reasonable estimate within a factor of $\sim 2$.  All cases in Fig.~\ref{inserr} scale as $\Omega^{3/2}$ (the respective fits are shown by lines).

In the remaining part of this Appendix we show that Eq.~(\ref{sigin2}) (with $\lambda$ being the chemical potential)  is a good approximation, i.e., that the covariance contribution to $\sigma^2_{\rm in}$ in (\ref{sigin2}) is small.

Eq.~(\ref{sesp}) holds for an arbitrary parameter $\zeta$ replacing $\lambda$. Since the number of particles $N$ is conserved, we have in general
 \begin{equation}
\label{sigin3}
\sigma^2_{\rm in} =  4\sum_i (\varepsilon_i-\zeta)^2 \sigma^2_{n_i} + A(\zeta) \;,
\end{equation}
 where
 \begin{equation}\label{A-lambda}
A(\zeta)= 4 \sum_{i\neq j} (\varepsilon_i - \zeta)(\varepsilon_j - \zeta) \mbox{cov}(n_i,n_j) \;,
\end{equation}
and $\mbox{cov}(n_i,n_j) = \overline{n_i n_j}-\bar n_i \bar n_j$ is the covariance of $n_i$ and $n_j$.

\begin{figure}[htb]
\includegraphics[width=\columnwidth]{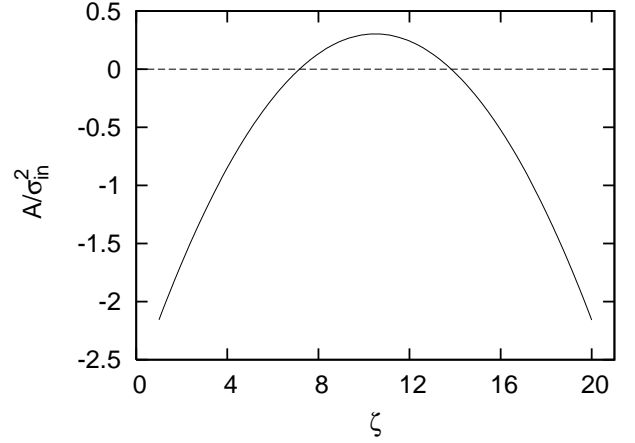}
\caption{The variation of the fractional error $A/\sigma_{\rm in}^2$ (see Eq. (\ref{A-lambda})) as a function of $\zeta$ for a uniform single-particle spectrum (level spacing of 1 MeV) with $\Omega=20$ at half filling and a constant pairing interaction $V_{ij}=0.3$ MeV.}
\label{zeta}
\end{figure}

The covariance contribution to (\ref{sigin3}) vanishes for values of $\zeta$ for which $A(\zeta)=0$, a quadratic equation in $\zeta$. Using particle number conservation $2\sum_j n_i = N = 2\sum_j \bar{n}_i$, we have
\begin{equation}\label{conserv}
n_i - \bar n_i = - \sum_{j \neq i} (n_j - \bar n_j) \;.
\end{equation}
Eq.~(\ref{conserv}) implies that
\begin{equation}
\sigma^2_{n_i} = -\sum_{j \neq i} \mbox{cov} (n_i,n_j) \;.\label{vc}
\end{equation}
Using Eq.~(\ref{vc}), we can express the coefficients of $\zeta^2$ and $\zeta$ in the quadratic equation $A(\zeta)=0$  in terms of the variances alone. We then find the following two solutions
\begin{equation}
\label{lampm}
\zeta_{\pm}= \zeta_0 \pm \sqrt{1 + \frac{\displaystyle \sum_{i\neq j} \varepsilon_i \varepsilon_j \mbox{cov}(n_i,n_j)}{\zeta_0^2 \sum_{i} \sigma_{n_i}^2}} \;,
\end{equation}
where
\begin{equation}
\label{lam0}
\zeta_0 = \frac{\displaystyle \sum_i \varepsilon_i \sigma^2_{n_i} }{\displaystyle \sum_i \sigma^2_{n_i}} = \frac{\displaystyle \sum_i \varepsilon_i \bar{n_i}(1-\bar{n_i}) }{\displaystyle \sum_i \bar{n_i}(1-\bar{n}_i)}
\end{equation}
is the midpoint between the two solutions.

As an example we show in Fig.~\ref{zeta} the quantity $A/\sigma_{\rm in}^2$ as a function of $\zeta$ for a uniform single-particle spectrum.
In general the zeros $\zeta_{\pm}$ of $A$ are not known without performing a full CSMC calculation.  However, we observe that $|A/\sigma^2_{\rm in}|$ is rather small in the region between $\zeta_-$ and $\zeta_+$ as compared with its typical value outside this region. Thus taking $\zeta \approx \zeta_0$  in Eq.~(\ref{sigin3}) and ignoring $A$ leads to a good approximation to $\sigma^2_{\rm in}$.

The sums on the r.h.s.~of Eq.~(\ref{lam0}) are dominated by those levels $i$ for which $\bar{n}_i$ is close to $1/2$, i.e., by levels in the vicinity of the chemical potential $\lambda$. Thus we expect $\zeta_0$ to be in proximity to the chemical potential. We can also estimate $\zeta_0$ directly from Eq.~(\ref{lam0}) using the BCS expressions for $\bar{n}_i$ in Eq~(\ref{nibar}).

\end{document}